# Out-of-time-ordered correlators in many-body localized systems

*Yichen Huang\*, Yong-Liang Zhang\*\*, and Xie Chen°*

In many-body localized systems, propagation of information forms a light cone that grows logarithmically with time. However, local changes in energy or other conserved quantities typically spread only within a finite distance. Is it possible to detect the logarithmic light cone generated by a local perturbation from the response of a local operator at a later time? We numerically calculate various correlators in the random-field Heisenberg chain. While the equilibrium retarded correlator $A(t=0)B(t>0)$ is not sensitive to the unbounded information propagation, the out-of-time-ordered correlator $A(t=0)B(t>0)A(t=0)B(t>0)$ can detect the logarithmic light cone. We relate out-of-time-ordered correlators to the Lieb-Robinson bound in many-body localized systems, and show how to detect the logarithmic light cone with retarded correlators in specially designed states. Furthermore, we study the temperature dependence of the logarithmic light cone using out-of-time-ordered correlators.

In the presence of disorder, localization can occur not only in single-particle systems [1], but also in interacting many-body systems [2–12]. The former is known as Anderson localization (AL), and the latter is called many-body localization (MBL). Neither AL nor MBL systems transfer energy, charge, or other local conserved quantities: Changes in energy or charge at position $x=0$ from equilibrium can spread and lead to changes in the corresponding quantity only within a finite distance $|x| < L_0$, where $L_0$ is the localization length.

A characteristic feature that distinguishes MBL from AL lies in the dynamics of entanglement after a global quench. Initialized in a random product state at time $t=0$, the half-chain entanglement entropy remains bounded in AL systems [13], but grows logarithmically with time in MBL systems [14–21]. In sharp contrast to the transport phenomena, the unbounded growth of entanglement in MBL systems suggests that information propagates throughout the system, although very slowly.

The propagation of information can be formulated by adapting the Lieb-Robinson (LR) bound [22–24] to the present context. In particular, it is manifested as the non-commutativity of a local operator $A$ at $x=0$ and $t=0$ with another local operator $B$ at position $x$ and evolved for some time $t$. In MBL systems, the operator norm of the commutator $[A(0,0), B(x,t)]$ is non-negligible inside a light cone whose radius is given by $|x| \sim \log|t|$, and decays exponentially with distance outside the light cone, i.e. [25],

$$\|[A(0,0), B(x,t)]\| \le Ce^{-(|x|-v_{LR}\log|t|)/\xi} \quad (1)$$

after averaging over disorder. Here, $B(x,t) = e^{iHt}B(x,0)e^{-iHt}$ is the time-evolved operator; $\|\cdot\|$ is the operator norm (the largest singular value); $C, v_{LR}, \xi$ are positive constants.

Is it possible to detect the logarithmic light cone (LLC) with equilibrium correlators of $A(0,0)$ and $B(x,t)$? Arguably the most straightforward approach is to measure the commutator in the LR bound (1) on equilibrium states, i.e., thermal states or eigenstates, using the Kubo formula in linear response theory:

$$\langle U^\dagger B(x,t)U\rangle - \langle B(x,t)\rangle = \tau\langle i[A(0,0), B(x,t)]\rangle + O(\tau^2), \quad (2)$$

where $U = e^{-iA\tau}$ with $\tau \ll 1$ is a local unitary perturbation if $A$ is Hermitian. The first and second terms on the left-hand side of (2) are the expectation values of $B$ in the presence and absence of the perturbation $U$, respectively. The difference is the effect of $U$ observed by measuring $B$. Note that $U$ and $B$ can be, but do not have to be, chosen as the operators for changing and measuring local conserved quantities, respectively. In this way, the effect of $U$ cannot be detected outside the localization length $L_0$.

---

\* Corresponding author, E-mail: ychuang@caltech.edu
\*\*Corresponding author, E-mail: ylzhang91@caltech.edu
° Corresponding author, E-mail: xiechen@caltech.edu
Department of Physics and Institute for Quantum Information and Matter, California Institute of Technology, Pasadena, California 91125, USA

Instead, we consider the out-of-time-ordered (OTO) correlator $A(0,0)B(x,t)A(0,0)B(x,t)$. Proposed long ago in the context of superconductivity [26], OTO correlators have received renewed interest in the study of the AdS/CFT correspondence, black holes, butterfly effect, quantum chaos, and scrambling [27–36]. Recently, several protocols have been proposed for measuring OTO correlators in experiments [37–39]. In MBL systems, we observe that OTO correlators can detect LLC. To understand this, we relate the OTO correlator to the Frobenius norm of $[A(0,0), B(x,t)]$, which is expected to exhibit similar scaling behavior as the operator norm in the LR bound (1). Furthermore, we study the temperature dependence of LLC using OTO correlators.

The observation that the expectation value of the commutator $[A(0,0), B(x,t)]$ with respect to thermal states or eigenstates cannot detect LLC described by the LR bound (1) raises the question of whether it is possible to saturate the bound (1) and thus detect LLC by measuring the commutator on some states using the Kubo formula (2). Of course, we can take the eigenstate of the commutator with the largest eigenvalue in magnitude, but such states are arguably not physical. Is it possible to detect LLC by measuring the commutator on physically motivated states? We provide numerical and analytical evidence that this is not the case for eigenstates of the Hamiltonian, any mixture of them, or random product states.

Finally, we compare AL with MBL. In AL systems, propagation of information forms a non-expanding light cone [40]. We show this explicitly using OTO correlators in the random-field $XX$ chain. The distinction between AL and MBL systems, i.e., non-expanding versus LLC, is also manifested in quantum revivals [41], modified spin-echo protocols [42], relaxation of local observables after a quantum quench, growth of connected correlators [19, 43], etc.

## 1 OTO correlators detect LLC

As a concrete model of MBL, we consider the spin-1/2 random-field Heisenberg chain and calculate its dynamics using exact diagonalization. The Hamiltonian is

$$H = \sum_{j=1}^{L-1} \left( \sigma_j^x \sigma_{j+1}^x + \sigma_j^y \sigma_{j+1}^y + \sigma_j^z \sigma_{j+1}^z \right) + \sum_{j=1}^{L} h_j \sigma_j^z, \quad (3)$$

where $\sigma_j^x, \sigma_j^y, \sigma_j^z$ are the Pauli matrices at the site $j$, and $h_j$'s are independent and identically distributed (i.i.d.) uniform random variables on the interval $[-h, h]$. This model is known to be in the MBL phase for $h \gtrsim 7$ [44–46]. We take $h = 16$ so that the localization length $L_0$ is small, and $L = 11$ unless otherwise stated. We perform a finite-size scaling analysis (data not shown for clarity) to ensure that finite-size effects are negligible.

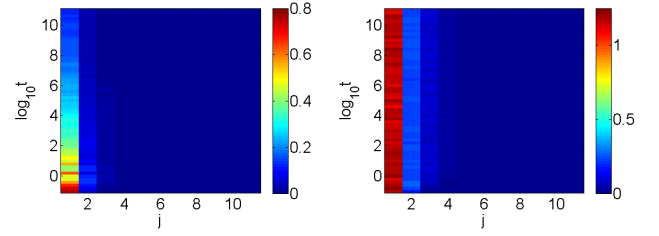

**Figure 1** Color images of $|\langle i[\sigma_1^x, \sigma_j^x(t)]\rangle_{\text{th}}|$ for $\beta = 0.1$ (left) and $|\langle i[\sigma_1^x, \sigma_j^x(t)]\rangle_{\text{eig}}|$ (right), averaged over 480 samples. LLC cannot be detected in this way.

Preparing the system in thermal states or eigenstates, the expectation value of an operator $\hat{O}$ is given by

$$\langle \hat{O} \rangle_{\text{th}} = \text{tr}(e^{-\beta H} \hat{O})/\text{tr}(e^{-\beta H}), \quad \langle \hat{O} \rangle_{\text{eig}} = \langle \psi | \hat{O} | \psi \rangle, \quad (4)$$

where $\beta = 1/T \geq 0$ is the inverse temperature, and $|\psi\rangle$ is a random eigenstate of $H$.

Figure 1 shows the expectation value of the commutator $[\sigma_1^x, \sigma_j^x(t)]$ with respect to thermal states and eigenstates as a function of $j = 1, 2, \ldots, L$ and $t$. Explained previously using the Kubo formula (2), this quantity measures the response at the site $j$ and time $t$ to a perturbation at the first site and $t = 0$. We see that the effect of perturbation spreads only within a small distance. We observe similar behavior (data not shown) for other choices of local operators in the commutator.

Let us consider the OTO operator $\sigma_1^x \sigma_j^x(t) \sigma_1^x \sigma_j^x(t)$. It is close to the identity operator if $[\sigma_1^x, \sigma_j^x(t)] \approx 0$, i.e., the site $j$ at time $t$ is outside the light cone generated at the first site and $t = 0$; its expectation value may deviate from 1 if the site $j$ is in the light cone. The top panels of Fig. 2 show this OTO correlator. We see LLC, in which the OTO correlator decays to zero. The bottom left panel shows the OTO correlator $\sigma_1^z \sigma_j^z(t) \sigma_1^z \sigma_j^z(t)$. We do not see a clear unbounded light cone. The bottom right panel shows $(\sigma_1^x + \sigma_1^z)(\sigma_j^x(t) + \sigma_j^z(t))(\sigma_1^x + \sigma_1^z)(\sigma_j^x(t) + \sigma_j^z(t))$. We see LLC, but in which the OTO correlator does not decay to zero.

The behavior of these OTO correlators can be understood from the local integrals of motion [20, 47] in MBL systems. The operators $\sigma_j^x, \sigma_j^z, \sigma_j^x + \sigma_j^z$ have vanishing, large, moderate overlap with the local integrals of motion. Therefore, the corresponding OTO correlator decays to zero, almost does not decay, decays to a finite value, respectively.

Why do OTO correlators behave differently from normal (i.e., equilibrium retarded) correlators? We have

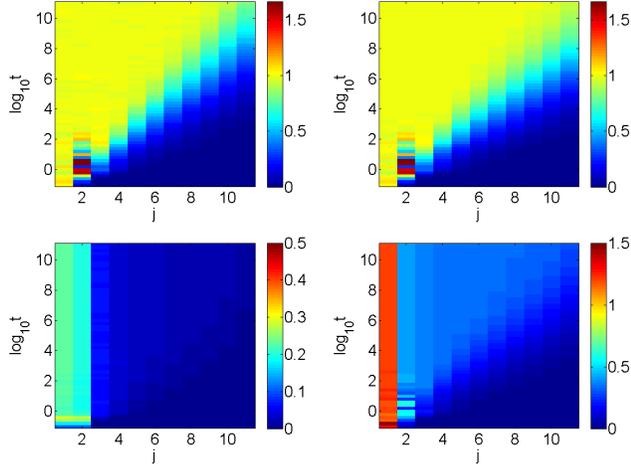

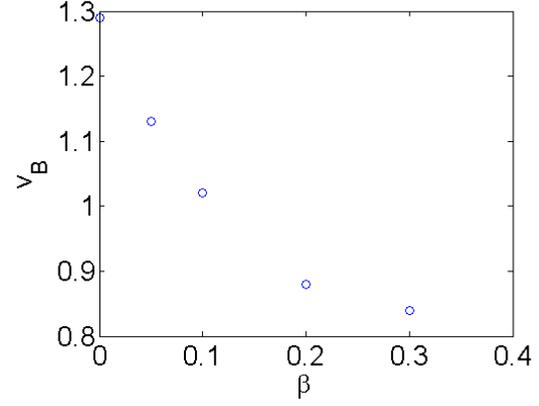

**Figure 2** Color images of $1 - \text{Re}\langle \sigma_1^x \sigma_j^x(t) \sigma_1^x \sigma_j^x(t)\rangle_{\text{th}}$ for $\beta = 0.1$ (top left), $1 - \text{Re}\langle \sigma_1^x \sigma_j^x(t) \sigma_1^x \sigma_j^x(t)\rangle_{\text{eig}}$ (top right), $1 - \text{Re}\langle \sigma_1^z \sigma_j^z(t) \sigma_1^z \sigma_j^z(t)\rangle_{\text{eig}}$ (bottom left), and $1 - \text{Re}\langle (\sigma_1^x + \sigma_1^z)(\sigma_j^x(t) + \sigma_j^z(t))(\sigma_1^x + \sigma_1^z)(\sigma_j^x(t) + \sigma_j^z(t))\rangle_{\text{eig}}/4$ (bottom right), averaged over 480 samples. We see LLC in all but the bottom left panels.

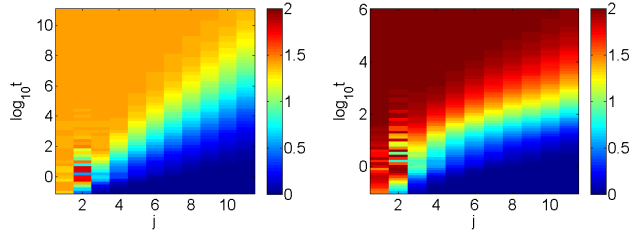

**Figure 3** Color images of $\|[\sigma_1^x, \sigma_j^x(t)]\|_F / 2^{L/2}$ (left) and $\|[\sigma_1^x, \sigma_j^x(t)]\|$ (right), averaged over 480 samples. The left panel is essentially the same (see Eq. 5) as the top right panel of Fig. 2. The right panel shows that the LR bound (1) is tight.

**Figure 4** LLC boundary $j \sim v_B \log_{10} t$ at various inverse temperatures $\beta$. The values of $v_B$ are obtained by solving (6) for $\epsilon = 1/2$. The top panels of Fig. 2 are the color images for $\beta = 0, 0.1$, and the color images for other values of $\beta$ are not shown. We see that $v_B$ decreases as $\beta$ increases.

Figure 3 shows the Frobenius and operator norms of the commutator. They exhibit similar scaling behavior in the sense of LLC in both panels. Thus, we have related OTO correlators to the LR bound (1).

We emphasize the difference between the expectation values of $i[\sigma_1^x, \sigma_j^x(t)]$ and $-[\sigma_1^x, \sigma_j^x(t)]^2$ shown in Figs. 1, 2, respectively. The observation that the latter detects LLC while the former does not can be understood as follows. The traceless Hermitian operator $i[\sigma_1^x, \sigma_j^x(t)]$ has both positive and negative eigenvalues, which may cancel themselves out upon taking the expectation value (with respect to either thermal states or eigenstates of the Hamiltonian). The eigenvalues of the positive semidefinite operator $-[\sigma_1^x, \sigma_j^x(t)]^2$ are nonnegative and contribute additively when taking the expectation value.

We now study the temperature dependence of LLC using OTO correlators (the temperature dependence of the linear light cone in homogeneous systems has been studied; see, e.g., [36]). To determine the LLC boundary $j \sim v_B \log_{10} t$, we choose a threshold $0 < \epsilon < 1$ and solve the relationship between $j$ and $t$ in the equation

$$1 - \text{Re}\langle \sigma_1^x \sigma_j^x(t) \sigma_1^x \sigma_j^x(t)\rangle_{\text{th}} = \epsilon. \tag{6}$$

We see from the top panels of Fig. 2 that $v_B$ depends on both $\beta$ and $\epsilon$. For fixed $\epsilon$, Fig. 4 shows that $v_B$ decreases as $\beta$ increases. This trend (faster information propagation at higher temperatures) was also found in some quantum field theories [36].

shown that normal correlators measure the response or spread of physical quantities like energy or charge after a perturbation from equilibrium. We now argue that OTO correlators describe the propagation of information.

The OTO correlator can be obtained by expanding the square of the commutator

$$-[\sigma_1^x, \sigma_j^x(t)]^2 = 2 - \sigma_1^x \sigma_j^x(t) \sigma_1^x \sigma_j^x(t) - \sigma_j^x(t) \sigma_1^x \sigma_j^x(t) \sigma_1^x$$
$$\Rightarrow 1 - \langle \sigma_1^x \sigma_j^x(t) \sigma_1^x \sigma_j^x(t)\rangle_{\text{th}} = \|[\sigma_1^x, \sigma_j^x(t)]\|_F^2 / 2^{L+1}, \tag{5}$$

where $\beta = 0$, and $\|A\|_F = \sqrt{\text{tr}(A^\dagger A)}$ is the Frobenius norm. In comparison, the LR bound (1) concerns $\|[\sigma_1^x, \sigma_j^x(t)]\|$.

The model (3) for $h \lesssim 7$ has two mobility edges separating delocalized eigenstates in the middle of the energy

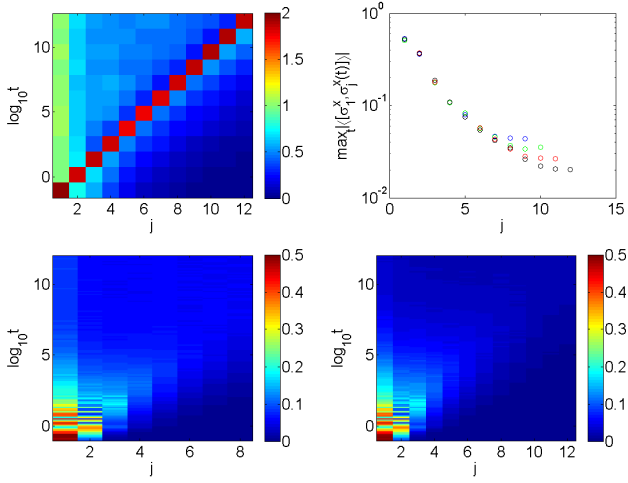

**Figure 5** (top left) Color image of $|\langle\psi_j|[\sigma_1^x,\sigma_j^x(t)]|\psi_j\rangle|$ for $L = 12$, averaged over 320 samples. Here, $|\psi_j\rangle$ is the eigenstate of $[\sigma_1^x,\sigma_j^x(t = 10^{0.3j+1})]$ with the largest eigenvalue in magnitude. (top right) The quantity $\max_t |\langle[\sigma_1^x,\sigma_j^x(t)]\rangle|$ on random product states for $L = 9$ (blue), 10 (green), 11 (red), 12 (black), averaged over 960 samples. It appears to decay exponentially with distance in the thermodynamic limit. (bottom) Color images of $|\langle[\sigma_1^x,\sigma_j^x(t)]\rangle|$ on random product states for $L = 8$ and $L = 12$.

spectrum from localized ones on the sides [45, 46]. For $h = 16$, (3) is deep in the MBL phase, and (almost) all eigenstates are localized. However, eigenstates in the middle of the spectrum might still be less localized than those on the sides. Therefore, we expect that $v_B$ decreases as $|\beta|$ increases. If a many-body localization-delocalization transition has mobility edges separating localized eigenstates in the middle of the spectrum from delocalized ones on the sides, we expect that $v_B$ increases as $|\beta|$ increases on the MBL side of the transition.

## 2 How to detect LLC with retarded correlators?

Is it possible to saturate the LR bound (1) and thus detect LLC by measuring the commutator on some states using the Kubo formula (2)? Of course, we can take the eigenstate of the commutator with the largest eigenvalue in magnitude. In the right panel of Fig. 3, the LLC boundary is roughly given by $t = 10^{0.3j+1}$. The top left panel of Fig. 5 shows $|\langle\psi_j|[\sigma_1^x,\sigma_j^x(t)]|\psi_j\rangle|$, where $|\psi_j\rangle$ is the eigenstate of $[\sigma_1^x,\sigma_j^x(t = 10^{0.3j+1})]$ with the largest eigenvalue in mag-

nitude. Thus, however far the measurement is from the perturbation, there is an initial state such that the effect of the perturbation can eventually be detected. In this sense, we do see LLC.

Since the eigenstates of the commutator are arguably not physical, is it possible to detect LLC by measuring the commutator on physically motivated states? As an attempt, we consider (random) product states, which are the initial states in most experiments on MBL [48–51]. The lower panels of Fig. 5 show the expectation value of the commutator with respect to random product states. We do not see a very clear unbounded light cone, but further analysis is necessary to draw a conclusion.

We consider the quantity $\max_t |\langle[\sigma_1^x,\sigma_j^x(t)]\rangle|$, which is the maximum signal that can ever be detected by measuring $\sigma_j^x$ (see Eq. 2). This quantity is not a function of $t$, and does not describe how the effect of a perturbation evolves in time. Rather, it detects whether the perturbation has long-range effects. Performing a finite-size scaling analysis, the top right panel of Fig. 5 provides evidence that in the thermodynamic limit, $\max_t |\langle[\sigma_1^x,\sigma_j^x(t)]\rangle|$ decays exponentially with $j$. Thus, a perturbation of random product states has only short-range effects at any time $t$, and LLC cannot be detected in this way. In the Appendix, we analytically calculate $|\langle[\sigma_1^x,\sigma_j^x(t)]\rangle|$ on some product states in the phenomenological model of MBL [20]. The analytical results agree qualitatively with the numerical results in Fig. 5.

## 3 OTO correlators in AL systems

In AL systems, information propagation forms a non-expanding light cone formulated by a strictly local LR bound [40]

$$\|[A(0,0),B(x,t)]\| \leq Ce^{-|x|/\xi}. \tag{7}$$

We now show this explicitly using OTO correlators. Thus, OTO correlators can distinguish AL from MBL.

As a model of AL, we consider the random-field $XX$ chain, which is equivalent to a model of free fermions hopping in a random potential. The Hamiltonian is

$$H = \sum_{j=1}^{L-1}\left(\sigma_j^x\sigma_{j+1}^x + \sigma_j^y\sigma_{j+1}^y\right) + \sum_{j=1}^{L} h_j\sigma_j^z, \tag{8}$$

where $h_j$'s are i.i.d. uniform random variables on the interval $[-h, h]$. This model is in the AL phase for any $h > 0$, and we still take $h = 16$. Figure 6 shows the OTO correlator $\sigma_1^x\sigma_j^x(t)\sigma_1^x\sigma_j^x(t)$ and the operator norm of the commutator $[\sigma_1^x,\sigma_j^x(t)]$. We see a non-expanding light cone.

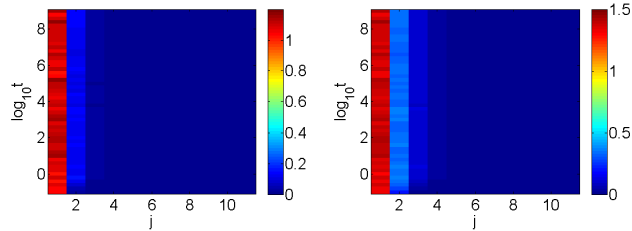

**Figure 6** Color images of $1 - \text{Re}\langle\sigma_1^x\sigma_j^x(t)\sigma_1^x\sigma_j^x(t)\rangle_{\text{eig}}$ (left) and $\|[\sigma_1^x,\sigma_j^x(t)]\|$ (right) in the AL system (8), averaged over 320 samples. We see a non-expanding light cone in both panels. This confirms (7).

## 4 Discussion and conclusions

We have shown that OTO correlators can detect LLC in MBL systems, and thus distinguish MBL from AL. Furthermore, we have studied the temperature dependence of LLC using OTO correlators. In the random-field Heisenberg chain (3), the LLC coefficient $v_B$ decreases as $\beta$ increases. However, it may be possible to construct models of MBL such that $v_B$ increases as $\beta$ increases.

In the linear response regime (2), we have studied whether a local perturbation of various initial states has long-range effects at a later time and thus can be used to detect LLC. We have shown that this is indeed the case for specially designed states, but may not be the case for thermal states, eigenstates of the Hamiltonian, or random product states. It would be interesting to study whether this is the case for other physically motivated states.

Beyond fully MBL systems, OTO correlators in a microcanonical ensemble can detect mobility edges. When the energy density of the ensemble is in the localized or delocalized region of the spectrum, we expect that OTO correlators demonstrate a logarithmic or power-law light cone, respectively. Furthermore, the linear light cone in diffusive non-integrable systems can be detected by measuring energy after a local perturbation of specially designed states in the linear response regime (2). To justify this, we have performed calculations (data now shown) as in the top left panel of Fig. 5 for the model in Ref. [52].

Recently, OTO correlators have been used extensively in quantum gravity to study chaos and scrambling of black holes via AdS/CFT duality. This paper shows that OTO correlators can provide insights into the dynamics of quantum many-body systems without a holographic dual.

## Note added

Shortly before and after this paper appeared on arXiv, we became aware of some related works. Ref. [53] studied LLC in MBL systems. Refs. [54–58] studied OTO correlators in MBL systems. Ref. [59] studied OTO correlators in marginal MBL systems.

**Acknowledgements.** We would like to thank Yoni BenTov, Manuel Endres, Yingfei Gu, Alexei Y. Kitaev, Joel E. Moore, and Rahul Nandkishore for inspiring discussions. In particular, Y.G. suggested studying the temperature dependence of LLC. Y.H. and X.C. acknowledge funding provided by the Institute for Quantum Information and Matter, an NSF Physics Frontiers Center (NSF Grant PHY-1125565) with support of the Gordon and Betty Moore Foundation (GBMF-2644). X.C. is also supported by the Walter Burke Institute for Theoretical Physics.

**Key words.** Anderson localization, Lieb-Robinson bound, many-body localization, out-of-time-ordered correlator.

## Appendix: Linear response for non-eigenstates

We compute analytically the linear response for a particular class of non-eigenstates using the phenomenological "l-bit" description of MBL [20]. The analytical result qualitatively supports our numerical observations in Fig. 5.

The Hamiltonian in the "l-bit" basis reads

$$H = \sum_i h_i \tau_i^z + \sum_{ij} J_{ij} \tau_i^z \tau_j^z + \cdots, \qquad (9)$$

where $J_{ij}$ decays exponentially with the distance between $i$ and $j$. For simplicity and without loss of generality, we ignore three- or more-body terms in the Hamiltonian.

We will compute the expectation value of $[\tau_1^x(t), \tau_{j_0}^x]$ with respect to product states in the $\tau^y$ basis. Note that the expectation value with respect to product states in the $\tau^z$ basis is clearly zero, reflecting the absence of transport in eigenstates of the Hamiltonian. By using product states in the $\tau^y$ basis, we have access to the off-diagonal elements of the commutator. However, we show that the linear response of such states still falls short of achieving the unbounded light cone.

The time evolution of $\tau_1^x$ is given by

$$e^{iHt}\tau_1^x e^{-iHt} = e^{ith_1\tau_1^z} e^{it\sum_j J_{1j}\tau_1^z\tau_j^z} \tau_1^x e^{-it\sum_j J_{1j}\tau_1^z\tau_j^z} e^{-ith_1\tau_1^z}$$

$$= e^{ith_1\tau_1^z} \times \prod_j \left(\cos(tJ_{1j}) + i\sin(tJ_{1j})\tau_1^z\tau_j^z\right) \times \tau_1^x$$

$$\times \prod_j \left(\cos(tJ_{1j}) - i\sin(tJ_{1j})\tau_1^z\tau_j^z\right) \times e^{-ith_1\tau_1^z}. \qquad (10)$$

Conjugating $\tau_1^x$ with $\prod_j (\cos(tJ_{1j}) + i\sin(tJ_{1j})\tau_1^z\tau_j^z)$ generates the following terms:

$$\tau_1^x,\ \tau_1^y\tau_j^z,\ \tau_1^x\tau_j^z\tau_k^z,\ \tau_1^y\tau_j^z\tau_k^z\tau_l^z,\ \ldots. \qquad (11)$$

Conjugating these terms with $e^{ith_1\tau_1^z}$ generates

$$\tau_1^{x\prime},\ \tau_1^{y\prime}\tau_j^z,\ \tau_1^{x\prime}\tau_j^z\tau_k^z,\ \tau_1^{y\prime}\tau_j^z\tau_k^z\tau_l^z,\ \ldots, \qquad (12)$$

where $\tau_1^{x\prime}$ and $\tau_1^{y\prime}$ are, respectively, the results of rotating $\tau_1^x$ and $\tau_1^y$ by $e^{ith_1\tau_1^z} = \cos(th_1) + i\sin(th_1)\tau_1^z$. The commutator of these terms with $\tau_{j_0}^x$ involves terms

$$\tau_1^{y\prime}\tau_{j_0}^y,\ \tau_1^{x\prime}\tau_{j_0}^y\tau_k^z,\ \tau_1^{y\prime}\tau_{j_0}^y\tau_k^z\tau_l^z,\ \ldots. \qquad (13)$$

The expectation value of these terms with respect to product states in the $\tau^y$ basis is nonzero only for the first term. The "bare" expectation value (up to $\pm 1$) is $\cos(2th_1)$ accompanied with a factor $\tan(2tJ_{1j_0})\prod_k \cos(2tJ_{1k})$. Putting together, the expectation value of the commutator with respect to product states of the "l-bits" in the $\tau^y$ basis is $\cos(2th_1)\tan(2tJ_{1j_0})\prod_k \cos(2tJ_{1k})$.

When $tJ_{1j_0} \ll 1$, the second factor is approximately $tJ_{1j_0}$ and thus the expectation value decays exponentially with $j_0$. One might think of this as outside the light cone. However, the expectation value on the boundary of the supposed light cone $j_c \sim \log t$ also decays with $j_c$, because $\cos(2tJ_{1k}) < 1$ when $k < j_c$ and $\cos(2tJ_{1k}) \sim 1$ when $k > j_c$. Therefore, the supposed light cone "faints out" with distance. This is exactly the behavior we see in the bottom panels of Fig. 5: A seemingly logarithmically growing light cone disappears with time. Due to the exponential decay of expectation values at light cone boundaries, the expectation value for any $j$ and $t$ can be upper bounded by an exponentially decaying function of $j$, indicating that the light cone should be considered finite. The maximum achievable expectation value for fixed $j$ decays exponentially with $j$, similar to that in the top right panel of Fig. 5.